\documentclass[preprint]{aastex}

\shorttitle{A Near IR Survey of NGC7538}
\shortauthors{Balog et al.}

\begin{document}

\title{A Near-Infrared (JHK) Survey of the Vicinity of the HII region NGC 7538:
Evidence for a Young Embedded Cluster\altaffilmark{1}}

\author{Z. Balog\altaffilmark{2}}
\affil{Dept of Optics and Quantum Electronics, University of Szeged, D\'om
t\'er 9, Szeged, H-6720 Hungary}
\email{balogz@titan.physx.u-szeged.hu}

\author{S. J. Kenyon\altaffilmark{3}}
\affil{Smithsonian Astrophysical Observatory, 60 Gardner St. Cambridge,
MA 02138 USA}
\email{skenyon@cfa.harvard.edu}

\author{E. A. Lada\altaffilmark{3}}
\affil{University of Florida, 211 Space Sci. Blgd., Gainesville, FL 32611, USA}

\author{M. Barsony\altaffilmark{3}}
\affil{San Francisco State University, 1600 Holloway Ave., San Francisco, CA 94132, USA and Space Science Institute, 4750 Walnut Street, Suite 205, Boulder, CO 80301, USA}

\author{J. Vinko \& A. Gaspar}
\affil{Dept of Optics and Quantum Electronics, University of Szeged, D\'om
t\'er 9, Szeged, H-6720 Hungary}

\altaffiltext{1}{Observations reported here were obtained at the MMT
Observatory, a joint facility of the Smithsonian Institution and the
University of Arizona.}
\altaffiltext{2}{former pre-doctoral fellow at Smithsonian Astrophysical
Observatory, 60 Gardner St. Cambridge, MA 02138 USA}
\altaffiltext{3}{Visiting Astronomer, Kitt Peak National Observatory, National
Optical Astronomy Observatory, which is operated by the Association of
Universities for Research in Astronomy, Inc. (AURA) under cooperative agreement
with the National Science Foundation.}

\begin{abstract}
We describe the results of two near infrared ($K$-band) imaging surveys and
a three color (JHK) survey of the vicinity of NGC~7538. The limiting
magnitudes are $K \simeq 16.5$ and $K \simeq 17.5$ mag for the $K$-band
surveys and $K \simeq 15$ mag for the JHK survey. We identify more
than 2000 and 9000 near-infrared (NIR) sources on the images of the
two $K$-band surveys and 786 NIR sources in the JHK survey.
From color-color diagrams, we derive a reddening law for background
stars and identify 238 stars with NIR excesses. Contour maps indicate
a high density peak coincident with a concentration of stars with
NIR excesses. We identify this peak as a young, embedded cluster and
confirm this result with the $K$-band luminosity function,
color histograms, and color-magnitude diagrams. The center of the cluster
is at $RA$ = 23:13:39.34, $DEC$ = 61:29:18.9. The cluster radius is $\sim$
3' $\sim$ 2.5 pc for an adopted distance, $ d \simeq$ 2.8 kpc. For
$d =$ 2.8 kpc, and reddening, $E_{J-K}$ = 0.55 mag, the slope of the
logarithmic $K$-band luminosity function (KLF) of the cluster,
$s \sim$ 0.32 $\pm 0.03$, agrees well with previous results for L1630
\citep[$s = 0.34$;][]{Lada91} and M17  \citep[$s = 0.26$;][]{Lada91b}.

\end{abstract}

\keywords{ open clusters and associations: individual (NGC~7538) --  stars:
formation -- stars: luminosity function -- stars: pre--main-sequence}

\section{Introduction}

In the last decade, large format infrared arrays have enabled direct
probes of the interiors of molecular clouds invisible at optical
wavelengths. The discovery of embedded clusters containing many newly
formed stars demonstrates the effectiveness of NIR imaging surveys
\citep[e.g.][]{Lada91,Lada93,Zinn93}. These studies show that young
clusters contribute significantly to star formation in our galaxy
\citep{Lada03}.

Young embedded clusters offer unique opportunities to study the stellar
initial mass function (IMF) and its variation in space and time. Dense
young clusters with ages of 10 million years or less probably have
nearly all of their original stellar population. Older clusters lose
members due to evolution and dynamical interactions \citep[e.g.][]{Frie95,
Meib02,Boil03,Boil03b,Bonn03}. The low mass members of young clusters are
brighter than at any other time of their evolution and are easier to
identify and to count than at later stages \citep{Lada03,Muen02,Muen03}.

NGC~7538 ($l$ = 111.5, $b$ = 0.8) is a visible HII region in the Perseus
spiral arm and is probably part of the Cas OB 2 complex. Lying at a distance
of $\sim$ 2.8 kpc \citep{Blit82, Camp88}, NGC~7538 is a site of
active star formation with
several luminous near-IR and far-IR sources ($L$ = 3000 - 30000 $L_\odot$;
$M$ = 10 - 20 $M_\odot$). \citet{Bloom98a}, \citet{Wern79}, and \citet{Wynn74}
catalogued 11 NIR sources and describe other properties of the region.

Although NGC~7538 is relatively well mapped in different wavelength
regimes from the optical to the sub-millimeter
\citep[e.g.][]{Momo01,Unge00,Yao00,Camp88}, there have been no large-scale
NIR surveys of the region since a short conference summary in \citet{McCa91}.
Here, we describe the first detailed analysis of a JHK imaging survey
in the vicinity of NGC~7538. The limiting magnitude of this survey,
$K \approx$ 15 mag, corresponds to stars with 2.0 $M_\odot$ for
$A_K$ = 1.1 mag and $d = 2.8$ kpc (adopted values from the literature).
This survey yields a robust detection of an embedded cluster
containing many stars with NIR excesses.  Our results complement the
deeper, but less extensive NIR survey of \cite{Ojha04}.

We describe the observations and data reduction in \S2,
analyze the photometric data in \S3, and derive a $K$-band
luminosity function (KLF) in \S4. We conclude in \S5 with
a brief summary.

\section{Observation and Data Reduction}

During 4-10 June 1993, we acquired JHK images with SQIID \citep{Elli93}
on the KPNO\footnote{Kitt Peak National Observatory, National
Optical Astronomy Observatory, which is operated by the Association of
Universities for Research in Astronomy, Inc. (AURA) under cooperative
agreement with the National Science Foundation.} 1.3m telescope.
The 12 pairs of dithered frames have a scale of 1.37''/pixel and
covered about 230 arcmin$^2$ centered at RA = 23:13:48 and
DEC = 61:27:36 (J2000). Small maps of two control fields are centered at
RA = 23:13:48; DEC = 61:43:30 (J2000) and RA = 23:15:54; DEC = 61:27:36
(J2000).
During 24-26 June 2002, we supplemented these data with additional
$K$-band observations with STELIRCAM\footnote{http://linmax.sao.arizona.edu/help
/FLWO/48/stelircam/index.html} mounted on the 48'' telescope at
Fred Lawrence Whipple Observatory on Mt. Hopkins. The STELIRCAM data
have a scale of 1.2''/pixel
and covered 270 arcmin$^2$ centered at RA = 23:13:41 and DEC = 61:31:39.
Standard stars of \citet{Elia82} were observed each night of both observing
runs. These data yield transformations into the CIT photometric system
\citep{Bars97}.  Finally, we acquired deeper K-band data on an engineering
run of
FLAMINGOS\footnote{http://www.astro.ufl.edu/~elston/flamingos/flamingos.html}
mounted on the 6.5m MMT in June 2001. The fields cover 836 arcmin$^2$
in a 5x5 mosaic centered on RA = 23:13:44 and DEC = +61:27:49 with
0.2''/pixel spatial resolution. Each element of the mosaic consists
of 5 dithered images.

To reduce the SQIID images, we used standard IRAF\footnote{IRAF is
distributed by the National Optical Astronomy Observatory, which is
operated by the Association of Universities for Research in Astronomy,
Inc. under contract to the National Science Foundation.} routines for
flat-fielding and sky-subtraction \citep[see][for a detailed description
of the reduction of SQIID frames]{Bars97}.  For the STELIRCAM data, we
used the STELIRCAM
pipeline\footnote{http://cfa-www.harvard.edu/ircam/} of Bill Wyatt provided
by the SAO Telescope Data Center. This software uses procedures similar
to those described in \citet{Bars97}. To combine the dithered images of each
FLAMINGOS field, we used the dither package in addition to the standard
routines for flat-fielding and sky subtraction.  Fig. 1 shows a JHK
composite image of the region based on SQIID data.

We used SExtractor \citep{Berti96} to select sources in both surveys.
After confirming each detection by careful visual examination of the frames,
we considered an IR source as a detection if it could be identified on
more than one STELIRCAM or FLAMINGOS image.  For SQIID data, we included
sources for the final sample if detected in all three bands.

In the following analysis, we consider the 786 stars detected in the SQIID
survey as the main sample. For the analysis of the density maps we also use
the 1047 STELIRCAM sources with $K<$15 as a second complete sample, the 2223
STELIRCAM sources with $K<$16.5 as a third sample, and the $\sim$ 9000
FLAMINGOS sources with $K<$17.5 as a fourth sample.

We used SExtractor to perform aperture photometry on all confirmed sources.
We used a small circular aperture (3 pixels for SQIID and 6 pixels for
STELIRCAM) for fainter cluster stars to reduce the background noise.
Large circular apertures for the standard stars (5 pixels and 12 pixels
respectively for SQIID and STELIRCAM data) yield a more accurate standard
transformation. To bring the measurements onto a common scale, we applied
an aperture correction derived from observations of relatively bright
stars on each frame. Due to changes in the point spread function and
seeing, the aperture correction changed slightly from frame to frame.
The typical value was 0.1 mag with rms = 0.02 mag.

For the FLAMINGOS frames we performed psf photometry using the IRAF/DAOPHOT
packages. We carefully selected the psf stars for each image and used a 2nd
order gaussian psf profile to account for different psf distortions. We
derive errors of $<$ 0.1 mag for stars with $K <$ 17.5.  Because these
fields were acquired through light cirrus, we used 2MASS observations to
transform FLAMINGOS data to a standard photometric system.

We derived separate transformations to place the SQIID and STELIRCAM data
on a standard photometric system.  For the 431 sources common to both
surveys,  the average difference in the $K$-band magnitude is
$K_{SQIID} - K_{STELIRCAM}$ = 0.08 mag $\pm$ 0.16 mag.  The offset
is small, occurs in all fields, and is independent of $K$.  Because all
of the SQIID frames were
acquired at much larger elevation than the STELIRCAM frames, telluric
extinction is the likely cause of the offset. The large dispersion
is also independent of $K$ and is probably due to source variability.
More than 60\% of the stars with large $K$-band differences also have
large NIR excess emission. NIR excesses and variability are
characteristic of pre-main sequence (PMS) stars
\citep[see e.g.,][and references therein]{Eiro02}.

Comparisons between SQIID and FLAMINGOS $K$-band magnitudes confirm
the measured offsets. We identified nearly all SQIID sources as single
objects on the FLAMINGOS images. After eliminating variable stars, the
average offset is $K_{SQIID} - K_{FLAMINGOS}$ = 0.02 $\pm$ 0.18 for
$K <$ 15.

We compared our results with previous IR photometry from the
literature \citep{Bloom98a,Wern79,Wynn74}. The average magnitude difference
between our photometry and published data is zero. All of the sources with
published NIR data are PMS stars. Thus, variability probably causes the
significant, $\sim$ 0.5 mag, dispersion in the average.

Comparisons with 2MASS also suggest a small magnitude offset.  We
identified more than 95\% of the stars detected with SQIID in the 2MASS
database. After eliminating $\sim$ 9\% of the sample with large
magnitude offsets ($>$ 0.40 mag), the mean offset between SQIID
and 2MASS data is 0.02 mag $\pm$ 0.13 mag.  We conclude that our
magnitude scale is robust and that 9\% or more of the NIR sources
are variable stars.

\section{Analysis of the photometric data}

H~II regions often contain one or more concentrations of young stars.
Because the extinction in these regions is usually large, the stellar
ionization sources are usually completely invisible at optical wavelengths.
NIR images often reveal embedded clusters containing at least several
O-B stars and many other lower mass stars. These embedded clusters are
among the youngest known stellar systems. From systematic studies of
individual star forming regions and surveys of molecular cloud complexes,
\citet{Lada03} summarized properties of more than 100 known embedded
clusters. Systematic examinations of the 2MASS catalogue suggest that there
are at least 300 embedded clusters within 10 kpc \citep{Bica03,Dutr03}.

The properties of embedded clusters vary considerably from region to
region. Most clusters are identified from a measurement of an overdensity
of stars relative to the local background. The typical star density is
between 10 and 100 stars $\rm pc^{-2}$, compared to projected background
levels of 2--10 stars $\rm pc^{-2}$. The sizes and total masses of the
clusters are 0.5--1.0 pc and 30--1000 M$_\odot$ respectively.  The
clusters usually suffer high and variable reddening due to the presence
of the dusty parent molecular cloud. Many of the members often show near-IR
excesses, which indicate the presence of a circumstellar disk
\citep[see][for references]{Lada03}.

The study of these concentrations provides data on the early stages of
star formation. The fraction of stars with IR excesses constrains the
disk destruction timescale and thus the timescale for planet formation
\citep{Lada02}.
The large number
of young stars with identical ages and distances in a cluster allows
statistical estimates for the luminosity and mass functions using
off-cluster control fields.  This information yields tests for
the nature and the universality of the IMF \citep{Lada03,Muen03}.

\subsection{Stellar density distribution}

To test for the presence of a cluster in NGC~7538, we analyzed the
stellar density distribution. We used the kernel method \citep{Silv86}
described in \citet{Gome93}, which allows the fairly sparse stellar
surface density to be smoothed. Briefly, the technique derives the
surface density $D$ at a point on the sky $(\alpha, \delta)$ from a
weighted average of the observed stellar density smoothed over a
length, $h$.  The kernel function $K$ provides the weighting:
\begin{equation}
D(\alpha,\delta) = {1 \over h^2} \sum_{i=1}^n K(\alpha,\alpha_i,\delta,
\delta_i)
\end{equation}
\noindent
and depends on the separation,
$r^2 = (\delta-\delta_i)^2+(\alpha -\alpha_i)^2 cos^2 \delta$, of each
star. \citet{Gome93} derived satisfactory results using a gaussian
kernel, which we adopted for simplicity

\begin{equation}
K(\alpha,\alpha_i,\delta,\delta_i) = {1 \over 2 \pi} e^{-r^2 \over 2 h^2}
\end{equation}

\noindent
We adopted a smoothing length $h$ = 1.8'' for the sparser SQIID data,
$h$ = 1.3'' for the denser STELIRCAM data and $h$ = 0.9'' for the high spatial
resolution FLAMINGOS data.

To analyze the stellar distribution in NGC 7538, we restricted the data
to the common area covered by the SQIID, STELIRCAM, and FLAMINGOS surveys.
We also divided the STELIRCAM data into two samples, a sample with the
same limiting $K$ magnitude as the SQIID data and another sample using all
of the STELIRCAM data. Fig. 2 shows the stellar density contours overlaid
on maps of stellar positions. For comparison, Fig. 3 shows the density
contours overlaid on a single SQIID frame and a 2MASS image of approximately
the same area.

All of the stellar density maps indicate several high density peaks in
NGC~7538.  The SQIID and the STELIRCAM samples show a concentration of
stars centered at
$RA$ = 23:13:39.34, $DEC$ = 61:29:18.9 (J2000) (SQIID; Fig 2, upper left
panel) and $RA$ = 23:13:38.93, $DEC$ = 61:28:51.6 (J2000) (STELIRCAM; Fig 2,
upper right and lower left panels). In the deeper FLAMINGOS data (Fig. 2 lower right panel), the center of the concentration is roughly 1 arcmin to the NW,
at $RA$ = 23:13:36.35 $DEC$ = 61:30:08.8 (J2000). These centroids agree
reasonably well with the peak derived from sources in the 2MASS point
source catalog, $RA$ = 23:13:38.84 $DEC$ = 61:29:01.05 (J2000), and with
the southeastern part of the optical nebula centered at
$RA$ = 23:13:30.21 $DEC$ = 61:30:10.5 (J2000) \citep{Wynn74, Camp88}.
These peak positions suggest a cluster radius of $\sim$ 3', $\sim$ 2.5pc
for $d$ = 2.8 kpc.

The positions of the peaks in stellar density follow a clear trend.
The peak in the shallow SQIID data is roughly 1 arcmin SE of the
main peak in the deep FLAMINGOS data. A second concentration in the
FLAMINGOS data coincides with the STELIRCAM peak around (1, -1) coordinates
in the lower right panel of Fig 2 and lies between the SQIID peak and
the main FLAMINGOS peak. These changes in the peak stellar density
follow a trend first noted by \citet{McCa91}, who identified three
concentrations along a SE-NW sequence.  \citet{McCa91} suggested that
the three condensations form an age sequence, with the oldest set of
stars in the NW and the youngest in the SE \citep[see also][]{Ojha04}.
However the trend of \citet{McCa91} refers to a larger scale.  We consider
this hypothesis below.

Despite the good agreement between the derived peaks in stellar density,
the morphology of the contours changes with limiting magnitude and spatial
resolution. The STELIRCAM data in the top right panel show an extra
concentration of stars south of the main concentration which is missing
in the SQIID data. This concentration is more pronounced in the deeper
STELIRCAM data (bottom left).  The deeper data also show a third
concentration of stars to the east of the main concentration.
With similar magnitude cuts, the 2MASS data show the same features.

The density contours derived from the deeper FLAMINGOS data suggest
two main concentrations of comparable stellar density. The increasing
density of the NW concentration suggests that it may be the densest
part of the region.

Based on our shallow imaging data, the NGC7538 cluster at
$RA$ = 23:13:39.34, $DEC$ = 61:29:18.9 (J2000) is relatively sparse
compared to other clusters.
The measured central surface number density is 10.3 $\rm arcmin^{-2}$ for
the SQIID data, 14.7 $ \rm arcmin^{-2}$ for STELIRCAM data with $ K<15$,
16.9 $\rm arcmin^{-2}$ for all STELIRCAM sources, and 97 $\rm arcmin^{-2}$
for FLAMINGOS sources. These correspond to projected stellar surface
densities of 15.5 (d / 2.8 kpc)$^{-2}$ pc$^{-2}$ (SQIID),
22.3 (d / 2.8 kpc)$^{-2}$ pc$^{-2}$ (STELIRCAM, K $<$ 15),
25.6 (d / 2.8 kpc)$^{-2}$ pc$^{-2}$ (complete STELIRCAM sample),
and 146.5 (d / 2.8 kpc)$^{-2}$ pc$^{-2}$ (FLAMINGOS sample).
The uncertain distance leads to a factor of 2--4 uncertainty in the
projected surface density \citep{Blit82, Camp88}.

Scaled to the same limiting absolute magnitude, the stellar density in
NGC~7538 is about an order of magnitude smaller than the density of the
core of the Trapezium cluster
(325 stars pc$^{-2}$ for SQIID and STELIRCAM data and 694 stars pc$^{-2}$
for FLAMINGOS data) \citep{Muen02}. Compared to other young embedded
clusters with densities of 10 and 100 stars pc$^{-2}$
\citep[see][for references]{Lada03}, the concentration in NGC~7538
is sparser than average from the SQIID and STELIRCAM surveys but still
above the lower limits of $\sim$ 10--15 stars pc$^{-2}$ for Gem1 and Gem4
\citep{Lada03}. However, the FLAMINGOS data reveal many faint stars and
resolved multiple sources invisible in shallower, lower resolution
observations.  Because the ratio of stellar densities between NGC~7538
and the Trapezium increases from $\sim$ 0.06 at $K \sim$ 15 to
$\sim$ 0.20 at $K \sim$ 17.5, deeper images may demonstrate that the
stellar density in NGC~7538 rivals that of the richest embedded clusters.

\subsection{Reddening Law}

Fig 2. shows several small concentrations of stars whose shape and
visibility depend on the magnitude limit and the wavelength of observation.
This structure could be caused by patchy extinction across a single embedded
cluster, several clusters observed at different depths in the nebula, or
the superposition of several clusters at different distances. To see which
of these or other possibilities is correct, we need to consider the reddening
law and the amount of reddening as a function of position across the cluster.

PMS stars often show a characteristic near-IR excess compared with reddened
main-sequence stars \citep{Keny87,Bert88,Lada92}. Because IR excess objects
generally fall outside the reddening band on the two-color ($J-H,~H-K$)
diagram, knowledge of the IR reddening law is necessary to distinguish
PMS stars from main sequence stars.

Using the technique described in \citet{Keny98b}, we used background
stars to derive the reddening law for the cloud. The method derives
$J-H$ and $H-K$ color excesses for each star projected onto the cloud
relative to sources in a relatively unreddened control field far from
the cloud and assumes the reddening for a particular star is the average
(or median) color excess. The slope of the reddening law is then defined
as the slope of the best-fitting line to the color excess measurements.

\citet{Keny98b} also define the reddening probability function, which is
the chance of measuring a pair of color excesses, to quantify errors in
the estimate for the slope of the reddening law.
This density function depends on the distribution of colors in
the off-field and on-field stars. To measure the color distribution,
\citet{Keny98b} used a kernel density estimator with a simple kernel
and smoothing length \citep[see][]{Silv86}.

Fig. 4 shows the extinction measurements for stars with $K<14$ mag
in the direction of NGC 7538. Stars without NIR excess emission
yield $E_{J-H}/E_{H-K} = 1.69 \pm 0.04$. This slope is in the range,
$E_{J-H}/E_{H-K}$ = 1.57--2.08, derived for other clouds using the
same technique. Previous results suggest that the slope of the NIR
reddening law correlates with the star formation activity, with
shallower slopes in more active star formation regions
\citep{Keny98b,Gome01,Racc02}. The slope derived for NGC 7538 places
it between $\rho$ Oph ($E_{J-H}/E_{H-K}$ = 1.57) and Cha I
($E_{J-H}/E_{H-K}$ = 1.80).

\subsection{Color-magnitude and color-color diagrams}

Together with the reddening law, the NIR color-color diagram
provides one way to identify PMS stars in an embedded cluster. PMS stars
often show NIR excesses due to the presence of a circumstellar disk
\citep[][and references therein]{Keny96} which places them to right
of the reddening band in the $(J-H,~H-K)$ diagram. Fitting the
zero-age main sequence (ZAMS) and isochrones for PMS stellar evolution
tracks on the color-magnitude ($J,~J-K)$ and ($K,~H-K)$) diagram yield
estimates for the cluster parameters like reddening and distance.

The left panel of Fig. 5 shows the color-color diagram for NGC~7538.
The reddening law from the previous section allows us to divide
the diagram into several parts. The reddening band contains about
60\% of the NIR sources.  Most of the remaining stars have varying
amounts of NIR excess emission,  with NIR colors similar to colors
predicted by disk models \citep{Adams87,Keny87,Keny96,Meye97,Chia97}.
However, there is a small group of 15 stars with NIR colors that
are inconsistent with disk models, $H-K \approx$ 1--1.5 and
$J-H \approx$ 0.
These stars could be planetary nebulae \citep{Whit85}
or B[e] stars \citep{Zick00}. Mid-infrared photometry or NIR
spectroscopy is needed to classify these sources.  Finally, roughly
5\% of the sample lies above the reddening band.  Unless these sources
are dominated by scattered light from the cloud \citep[see][]{Whitn97},
the colors of these sources are unphysical. Because most of the sources
are relatively faint ($K>14$), we suspect that photometric error or
nebular contamination is responsible for the colors of these sources
(see below).

The stars in the reddening band show that the reddening in the field
of NGC~7538 is variable.  The observed color excesses range from
$E(H-K) \simeq 0.0$ to $E(H-K) \simeq 1.0$. For a normal reddening
law, these excesses yield a range in visual extinction from
$A_V \simeq 0$ mag to $A_V \simeq 16.5$ mag \citep{Bess88}.
Stars with modest excesses $E(H-K) \lesssim 0.2$ are scattered
randomly across the field and have colors similar to stars in the
control field. Most of these sources are probably nearby, with
distances of 1--2 kpc or less \citep[e.g.,][]{Neck80, Hakk97}.
The highly reddened stars are concentrated near the optical nebula,
in the main SQIID concentration of NIR sources described above.  Most
stars with colors above the reddening band or with colors similar
to planetary nebulae or B[e] stars are in the brightest parts of
the nebula, suggesting that direct or scattered light from the H II
region might contaminate the measured colors.

Fig. 2 shows that peaks in the stellar density distributions in the
SQIID and STELIRCAM datasets lie within a dense concentration of
stars in the deeper FLAMINGOS image. This concentration is roughly
1' SE of the main FLAMINGOS peak. More than half (56\%) of the stars
in this concentration have an IR excess, compared to 37$\%$ for the
total sample. Roughly 34$\%$ of all IR excess stars lie in this
concentration, which covers about 8$\%$ of the observed field. This
result provides further evidence for a young cluster in NGC~7538.

To test whether the NIR excess sources are clustered, we performed a two
dimensional KS-test \citep{Press93}. For simplicity, we defined a circular
area with radius of 0.05 deg centered on the density maximum (\S3.1) and
compared the
distribution of colors for stars within the circle to the colors of stars
outside the circle. We also checked whether sources with and without IR
excess are randomly distributed in the concentration (within the circle
with 0.05 deg radius). The probability that the two groups of sources are
selected randomly from the same parent distribution is smaller than $\sim
5 \times10^{-3}$ for both cases. We conclude that the NIR excess sources
preferentially lie in the cluster and that they are not from the same
population of stars as the stars without IR excess.

To test the conclusions of \citet{McCa91}, the right panel of Fig. 5
shows colors for three spatial regions in NGC~7538.  We identified 30
stars in region 1 of \citet{McCa91}. All of
these are heavily reddened.  In our data, regions 2 (131 stars including
IRS 1-3) and 3 (138 stars) have comparable numbers of stars. The
stars in region 2 are more heavily reddened than those in region 3 but
less heavily reddened than stars in region 1. On average, the stars in all
three regions are more heavily reddened than stars in the control field.
Thus the sequence in stellar density forms a sequence in reddening,
with the most heavily reddened stars in the SE portion of the nebula
and the least reddened stars in the NW portion \citep[see also][]{Ojha04}.

The IR color-magnitude diagram (CMD) also provides information on the
stellar population of the cluster. Comparisons between observations
and model isochrones yield approximate ages for stars in the cluster,
another measure of the reddening, and confirmation of the relative
numbers of stars with and without NIR excess.  The observations can
also provide tests of model isochrones, especially for the youngest
stars where the physics of stellar formation and evolution remains
uncertain \citep{Sies00,Bara98,Bara03,Burr97,Dant94}.

Fig. 6 and 7 show the color magnitude diagrams (Fig. 6: $K,~H-K$, Fig. 7.
$J,~J-K$) of NGC~7538. The left panels of each figure show all stars
in the sample; the right panels show only stars in the concentration
(see Fig. 5 for the meaning of the symbols). Stars with low reddening
are close to the
ZAMS in each figure.  For $E(J-K) =$ 0.55 and $E(H-K) =$ 0.19, the model
ZAMS indicates a distance of 2.1 kpc for these stars. This result
strengthens the conclusion that these stars are foreground to the
cluster.

Stars with NIR excess occupy different regions of the CMDs. Although
some stars lie close to the ZAMS or the 1 Myr isochrone with small
reddening, $E(J-K) =$ 0.50 and $E(H-K) =$ 0.19, most stars require
additional reddening. For a modest IR excess, $\sim$
0.2--0.5 mag at K, most NIR excess stars require $E(H-K) \approx$
0.4--0.7 and $E(J-K) \approx$ 1.2--1.8 to fall near the 1 Myr
isochrone of \citet{Sies00}.  For these isochrones, the most heavily
reddened stars are the youngest.  Stars with the largest NIR excesses
are also younger than stars with smaller NIR excess. However, the age
differences are small and require accurate spectral types for confirmation.

Tests with isochrones derived by other groups yield similar results.
For the \citet[][and references therein]{Dant97} and the
\citet{Pall99} tracks, most of the heavily reddened stars lie
between the ZAMS and 1--2 Myr isochrones \citep[see also][]{Ojha04}.
Stars with NIR excess
require $E(H-K) \approx$ 0.3--0.8 and $E(J-K) \approx$ 1--2, in
agreement with results for the \citet{Sies00} tracks. For these
young, relatively massive stars, the tracks are uncertain.  Thus, our age
estimates should be taken as an indication that the most heavily
reddened stars and stars with NIR excesses are young stars that
have not yet arrived on the ZAMS.

The PN or B[e] star candidates cluster around the same areas in each CMD.
These stars do not obviously fit the ZAMS or a pre-main sequence isochrone.
The stars above the reddening band in Fig. 5. are roughly randomly
distributed on both CMDs.  In both cases, flux from the nebula or
photometric errors probably displace these stars from the ZAMS or the
1 Myr isochrone.

To estimate a lower limit to the distance and the amount of extinction
to the concentration,
we fit the ZAMS of \citet{Sies00} to the blue edge of distribution of stars
in the $(J-K)-K$ diagram.
This exercise yields $E(J-K)$ = 0.55 $\pm$ 0.05 and a distance modulus of
11.6 $\pm$ 0.2 mag, which corresponds to a distance of about 2.1 kpc.
This distance agrees with the distance derived for the foreground stars
and is somewhat lower than the commonly accepted $d$ = 2.8 $\pm$ 0.9
\citep{Blit82, Camp88} for NGC~7538.  NIR spectra of sources in the
concentration would provide better estimates for the distance and reddening.

\section{Luminosity and color functions}

The cluster CMDs indicate that most of the young stars with NIR excess
fall close to a 1 Myr isochrone, with extinction and distance
appropriate for an embedded cluster within the NGC~7538 nebula.
Although we confirm the \citet{McCa91} identification of three distinct
concentrations of stars within the nebula, we do not confirm the
proposed range in stellar ages.  In our data, the large dispersion in
the reddening for stars in these groups precludes assigning accurate
ages. Because the shape of the luminosity function depends on the
age of a cluster \citep{Muen03}, we now consider whether we can
distinguish the ages of stars in the concentrations through analysis
of the luminosity function.

There are several ways to estimate the IMF from infrared observations.
Because stars in most clusters have measured colors but not spectra,
luminosity functions (LFs) or CMDs allow better statistical comparison
 between data and models than H-R diagrams (HRDs).
However, the evolutionary tracks used in model LFs are uncertain,
which leads to less precision compared to model HRDs. Color magnitude
diagrams enhance the precision by replacing effective temperature with
NIR colors, but extinction, NIR excess emission, and starspots complicate
assigning appropriate dereddened colors for pre-main sequence stars
\citep{Keny95}.

Among others, \citet{Lada91,Lada93b} used cumulative logarithmic KLFs to
estimate the slope of the IMF for several embedded clusters, including
NGC~2023, NGC~2024, NGC~2068 and NGC~2071 in the L1630 molecular cloud and
NGC~2264. More recently \citet{Muen02,Muen03} carried out a similar
analysis for IC 348 and the Trapezium.  To put NGC~7538 in context with
these studies, we analyze the KLF and then compare our results with
results for other regions.

Fig. 8 shows the $K$-band luminosity function of NGC~7538.
In the left panel the hollow histogram is the KLF of the concentration
(within a 0.05 deg circle around the peak stellar density). The filled
histogram is the KLF for the remaining sources. The right panel shows the
histogram of the two larger subgroups (\citet{McCa91}'s regions 2 and 3)
together with the histogram of sources outside the concentration. Due to
the small number of sources, the bins are relatively wide, 0.5 mag. We also
exclude sources from region 1 of \citet{McCa91}. The small number of stars
detected in this region precludes reliable statistics for the KLF.

The KLF has several main features which are observable in both panels.
The concentration has a larger
density of sources than the surroundings (\S3.1). The peak of the
KLF for the concentration is roughly 0.5 mag fainter than the KLF for the
surroundings. The shapes of the two histograms are also different.
Outside the concentration, there is a rapid rise before the peak.  Inside
the concentration, there is an excess of bright stars and the rise in the
histogram is more gradual. In the right panel, the peak of the population
in region 3 is shifted to fainter magnitudes and has an excess of fainter
stars relative to the rest of the condensation.  \citet{Muen02b} showed
that the KLFs of older clusters peak at fainter $K$ magnitudes.
Because region 3 has smaller extinction than the rest of the condensation,
the shift of the peak of the histogram thus strengthens the hypothesis of
\citet{McCa91} that region 3 contains older stars than region 2.

To test whether extinction is the most likely cause for the shift in the KLF
between the concentration and the remaining part of the area, we examined
the color distributions (Fig. 9 left panel). The stars
outside the concentration show a nearly Gaussian distribution around
one peak ($J-K \simeq 0.8-0.9$). The full-width at half maximum of
the gaussian is $\simeq 0.7$. The number of sources decreases sharply towards
larger J-K. In contrast, the concentration has a double-peaked histogram.
One group of stars has a distribution nearly identical to the off-cluster
histogram; the other group contains stars roughly 0.7-0.8 mag redder.
The dispersion in the red peak is roughly twice as large as the dispersion
in the blue peak.

In both histograms, foreground stars produce the blue peak. The density
of stars in front of the concentration and off the concentration is
roughly equal.  The peaks fall in the range occupied by MS stars with
masses exceeding 1 $M_{\odot}$ ($J-K \approx$ -0.18 to 0.51
\citep{Sies00}) and reddening, $E(J-K)$ = 0.55, smaller than the reddening
in the cluster (\S3.3).  The dispersion in the color distribution is
close to the range in NIR reddening expected from the optical extinction
maps \citep{Neck80, Hakk97}.

Extra reddening and near-IR emission emission produce the second peak
in the histogram for the concentration.  From the $E(H-K)$ values of
\S3.3 and our reddening law, $E(J-K)$ from reddening is 0.1--2.4. The
contribution of the intrinsic IR excess from a circumstellar disk is 
0.35--0.75 \citep{Keny96}. The combination of reddening and near-IR 
excess yields a total color excess of $\simeq$ 0.45--3.15 in J-K, 
consistent with the extent of the second peak in Fig. 9.

In summary, we can unambiguously confirm a sequence in reddening for the three
condensations of \citet{McCa91} but the CMD does not yield a clear conclusion
regarding an age sequence. The 1 Myr isochrone and ZAMS
are too close together to draw a meaningful conclusion. Deeper images,
which detect 0.1-0.5 Msun stars, might be able to distinguish the ages
of the regions. The KLF suggests an age sequence, but we cannot be certain.

The right panel of Fig. 9 provides a consistency check on our claim
that region 2 is more heavily reddened than region 3.  The peak of the
color histogram for region 2 is $\sim$ 0.8 mag redder than the peak of
region 3, confirming that this region is more heavily reddened.

The embedded cluster accounts for the breadth of the reddened histogram
in Fig. 9. Off the cluster, we see a modest range in galactic reddening.
On the cluster, we see galactic reddening plus a range in reddening
through the molecular cloud to the cluster. The separation of the
two peaks in the histogram implies an extra extinction of
$A_{K_{ext}} \simeq 0.5$ mag \citep{Bess88}.  The broad red tail of
the red histogram represents reddening in and beyond the cluster.
To demonstrate this conclusion, Fig. 10 plots the J-band LF for
(i) off the cluster, (ii) on the cluster but outside the main concentration,
and (iii) on the main concentration. The reddening clearly increases
along this sequence.

To analyze the cumulative KLF, we consider a complete sample of the
NIR survey. In the top panels of Fig. 11, the KLF contains stars with
$K <$ 15.5, which corresponds to stars of $ M >$ 1.15 M$_{\odot}$ at the
adopted cluster distance of 2.8 kpc.  From the 2MASS, SQIID, and STELIRCAM
data, we estimate a completeness limit of K $<$ 14.5, corresponding to stars
with masses of at least 1.55 M$_{\odot}$.  This limit is shallower than KLFs
for
most nearby clusters, which often reach below 1 M$_{\odot}$
\citep[see][]{Lada03}.

To derive the slope of the KLF for our complete sample, we separated the stars
into three main groups, (i) the concentration, (ii) the off-cluster field,
and (iii) the on-cluster field around the concentration. We further divided
the concentration into three groups according to \citet{McCa91}.
We counted the stars in each region and normalized this result. Subtracting
the normalized KLF of the off-cluster fields from the concentration
 yields the difference, log$_{10}$ ($KLF_c - KLF_o$), as a function
of K magnitude. A straight line fit to this function results in a slope,
$s$ = 0.29 $\pm$ 0.02. Fitting the logarithmic difference between the
concentration and the on-cluster field yields $s$ = 0.34 $\pm$ 0.02.
To limit errors in the slope of the KLF from small number statistics,
we limited this analysis to stars with $K$ = 11--14.5 mag.

To check whether the cumulative KLFs of different regions yield different
results we repeated the analysis for regions 2 and 3.  Despite the fact that
the KLFs for the two regions peak at different $K$ magnitudes, the derived
slopes for the two regions are indistinguishable from those for the
complete condensation (see Fig. 11).  Both results agree with estimates
of \citet{Lada91,Lada91b}. The similarity in the slopes supports the
hypothesis of \citet{Muen02b} for young clusters that the KLFs for
regions with large age spreads ($\delta \tau \sim$ 5 Myr) are hard to
distinguish from KLFs of regions where all the stars are the same age.

As a final check on the slope of the KLF, we repeated this analysis
for the deep FLAMINGOS data.  Here, we limited the analysis to stars
with $K$ = 11--16 mag. For all stars in the concentration, we derive
$s$ = 0.35 $\pm$ 0.01.  Dividing the area into the three regions
of \citet{McCa91} and all other stars, we derive $s$ = 0.25 $\pm$ 0.02
(region 1), $s$ = 0.35 $\pm$ 0.02 (region 2), and $s$ = 0.37 $\pm$ 0.02
(region 3). The change in $s$ from region 1 to regions 2 and 3 is real
and an indication that region 1 contains the youngest stars in NGC~7538.
\citet{Ojha04} derive similar results from their deeper survey of a
smaller area centered on the cluster.

If we assume a power law IMF and coeval star formation, we can derive
a crude estimate for the slope of the IMF from the slope of the
cumulative KLF. We define the IMF and the mass-luminosity relation as power
laws.

\begin{equation}
dN(log~ m_{\star}) \propto m_{\star}^{-\alpha} d log~m_{\star}
\end{equation}

\begin{equation}
L_K \propto m_{\star}^{\beta}
\end{equation}
\noindent
where $\alpha$ and $\beta$ are spectral indices of the IMF and the mass
luminosity relation, respectively \citep{Lada93b}. For O-F type stars
$\beta$ = 2.0 \citep{Lada93b}.
The slope of the logarithmic KLF is then \citep{Bloom98b}
\begin{equation}
s = {\alpha \over {2.5\beta}} .
\end{equation}

\citet{Bloom98b} also shows that the slope of the logarithmic KLF is 
equal to the slope of the cumulative KLF.

The average slope of the KLF ($s$ = 0.32) implies $\alpha \simeq 1.58$. This
$\alpha$ value is much steeper than the Salpeter value. Smaller $\beta$ would
result in a significantly shallower slope of the IMF.

With $\alpha$ known, the total mass in the concentration follows from
an integral over the IMF. To compare our result with estimates for other
young clusters we chose the  mass range of 1--120 $M_{\odot}$ for
the integration. For this mass range
the concentration contains $\sim$ 790 $M_{\odot}$.  This mass is
comparable to the mass in the Orion Nebula cluster and the Pleiades
($\sim$ 450 $M_{\odot}$; \citep{Slez02}) and 4--5 times smaller
than the mass in h (3700 $M_{\odot}$) or $\chi$ (2800 $M_{\odot}$)
Persei \citep{Slez02}.

\section{Conclusions}

We conducted two independent NIR surveys of the vicinity of the  galactic
H~II region  NGC~7538.  A shallow map of 230 arcmin$^2$ in JHK contains
$\sim$ 600
NIR sources with $K <$ 14; deeper K band imaging has $\sim$ 2000 NIR
sources with $K <$ 15.5 and 9000 sources with $K <$ 17.5. Our analysis
of these surveys yields the following results.

\begin{itemize}

\item In all surveys, the stellar density distribution is sharply
peaked at $RA$ = 23:13:39.37, $DEC$ = 61:29:13.01.  The peak stellar
density is within the range observed in other young clusters. The
deeper imaging data suggests the cluster has two or three main
concentrations.  2MASS data confirm this point.

\item We use the $J-H$, $H-K$ two-color diagram to derive the slope
of the reddening law, $E_{J-H}/E_{H-K}$ = 1.69. The slope of
the reddening law agrees with results for other star forming
regions. Stars within the main peak in the stellar density are
much more heavily reddened than stars in a control field well
off the peak.

\item The two-color diagram and the stellar density map indicate
that the cluster contains many young stars with NIR excesses.
The fraction of cluster stars with NIR excess is $\sim$ 30\%.
Cluster stars are more heavily reddened than stars in the
surrounding area. The NIR excess source fraction is somewhat
lower than the expected value of 80\%-50\% for a 1-3 Myr old
cluster. A fraction of 30\% is more typical of 4 Myr old clusters
\citep{Lada02} and indicates an age closer to $\sim$ 4 Myr for
stars in the SQIID sample. However, we note that fainter stars
in the region appear younger, suggesting that deeper JHK surveys
will detect stars with ages closer to $\sim$ 1 Myr \citep{Ojha04}.

\item Fits of the \citet{Sies00} isochrones to the blue edge of
the color magnitude diagram yield lower limits for the cluster
reddening $E_{J-K} \ge 0.55$, and the distance, $d \ge 2.1$ kpc.
These values agree with the previous estimates \citep[eg.][]{Blit82, Camp88}.
On the NIR CMD, most stars in the concentrations fall between a 1 Myr old
isochrone
and the ZAMS. However, the large spread in reddening among these
stars prevents a robust estimate of their age.

\item The slope of the cumulative logarithmic KLF of
the cluster is $s$ = 0.32 $\pm$ 0.03.  This slope agrees with
results for other star forming regions \citep{Lada91,Lada91b}.
The slope of the KLF for region 1 of \cite{McCa91} suggests that
stars in this region are much younger than stars in the rest of
the cluster.  The KLF does not yield an obvious difference in
age between regions 2 and 3 of \cite{McCa91}.

\end{itemize}

Finally, deeper multicolor imaging data and spectroscopic observations
with a large aperture telescope would allow improved constraints on our
derived parameters.  Deeper multicolor imaging would place better limits
on the stellar density in regions 1--3 and provide better estimates for
the variation in extinction from region to region.  NIR spectroscopy
would yield spectral types and allow the construction of a robust HRD.
Comparison with modern isochrones would provide more accurate ages and
a test of the apparent change in mean stellar ages from region 1 to
regions 2--3.

\acknowledgments
The authors would like to express their thanks to the anonymous referee for
the very thorough, critical and helpful comments and suggestions.
This work was
 supported by the Hungarian OTKA grants T034615, F043203, T042509 and the SAO
Predoctoral Fellowship program.

\clearpage

\begin{figure}[ht]
\begin{center}
\leavevmode
%\plotone{Balog.fig1.eps}
\caption{False color image of NGC~7538 (red=K, green=H, blue=J). The images covers a region of 13' x 11' with N to the top and E to the left. The pixel scle is 1.37''/pixel}
\end{center}
\end{figure}

\begin{figure}[ht]
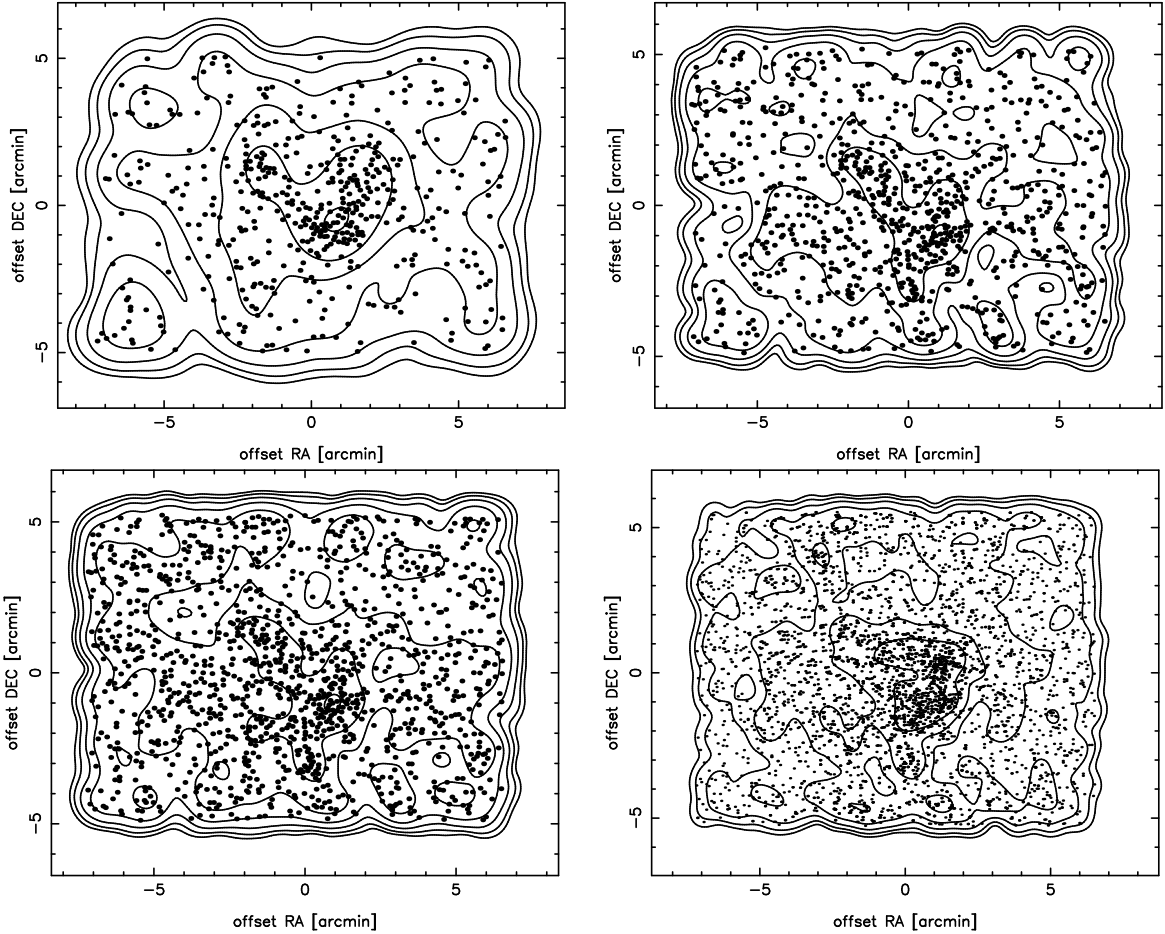

\begin{center}
\leavevmode
\plottwo{Balog.fig2a.eps}{Balog.fig2b.eps}
\plottwo{Balog.fig2c.eps}{Balog.fig2d.eps}
\caption{Smoothed stellar density contours at K for SQIID (top left panel), shallow STELIRCAM data (top right panel), deeper STELIRCAM data (bottom left panel) and FLAMINGOS (bottom right panel). The orientation of the plots is the same as in Fig. 1. Regions 1, 2 and 3 of \citet{McCa91} are at (-1,3), (1,0) and (2,1) respectively.}
\end{center}
\end{figure}

\begin{figure}[ht]
\begin{center}
\leavevmode
%\plottwo{Balog.fig3a.eps}{Balog.fig3b.eps}
\caption{The stellar density contours overlaid on one of the SQIID (left panel) and 2MASS (right panel) images. The orientation is the same as in Fig. 1, the FOV $\simeq$ 5' X 5' in both panels.}
\end{center}
\end{figure}

\begin{figure}[ht]
\begin{center}
\leavevmode
\plotone{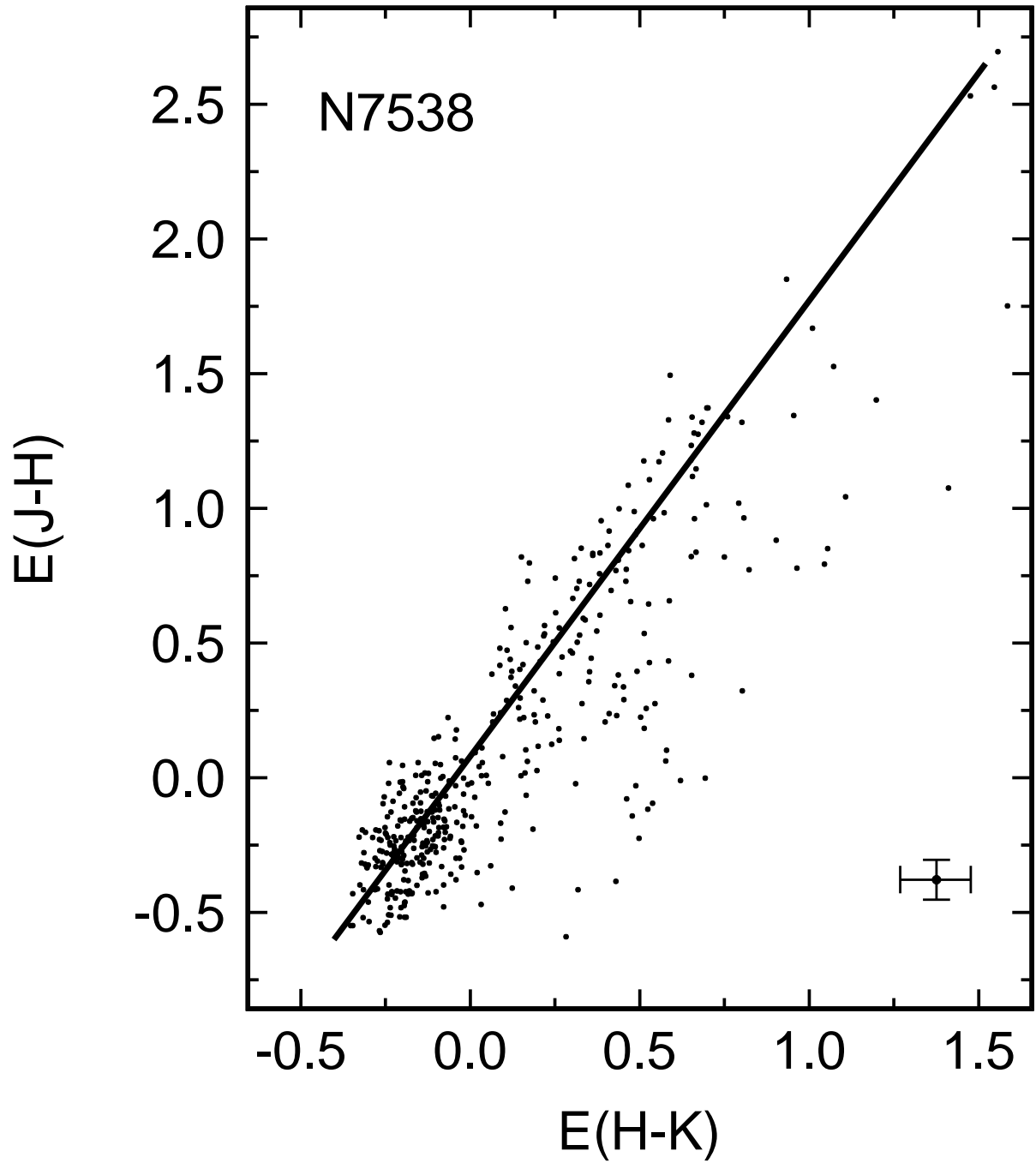}
\caption{Reddening law for NGC~7538.}
\end{center}
\end{figure}

\begin{figure}[ht]
\begin{center}
\leavevmode
\plottwo{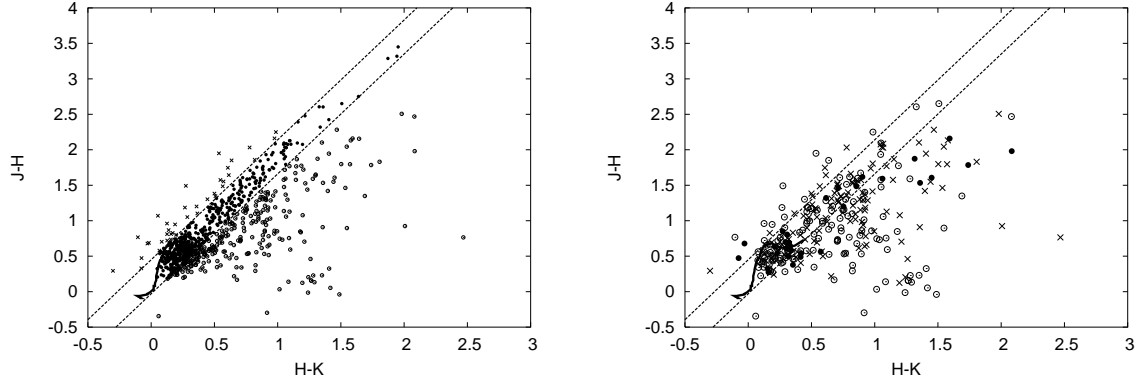}{Balog.fig5b.eps}
\caption{The color-color diagram of the detected IR sources. Left panel: all stars of the survey. Open circles: stars with infrared excess, filled circles: main sequence stars, x-es: stars above the reddening band. Right panel: the area divided into 3 regions according to \citet{McCa91}. Filled circles: region 1, x-es: region 2, open circles: region 3. See the definition of the regions in the text. Solid lines denotes the ZAMS of \citet{Sies00}; dashed lines represent the reddening band.}
\end{center}
\end{figure}

\begin{figure}[ht]
\begin{center}
\leavevmode
\plottwo{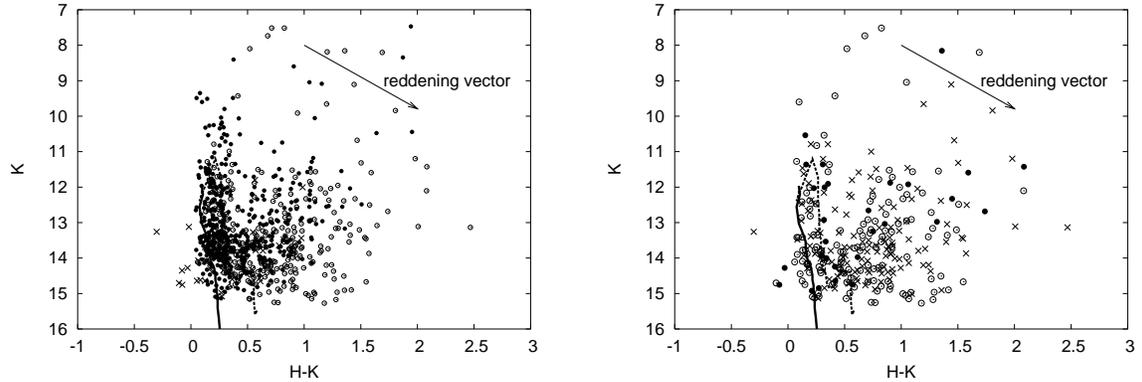}{Balog.fig6b.eps}
\caption{H-K vs K diagrams of the detected IR sources. Left panel: all stars. Right panel: stars divided into the \citet{McCa91} regions. See Fig. 5 for the meaning of the symbols. Solid line denotes the ZAMS of \citet{Sies00}; dashed line shows the 1 Myr isochrone.}
\end{center}
\end{figure}

\begin{figure}[ht]
\begin{center}
\leavevmode
\plottwo{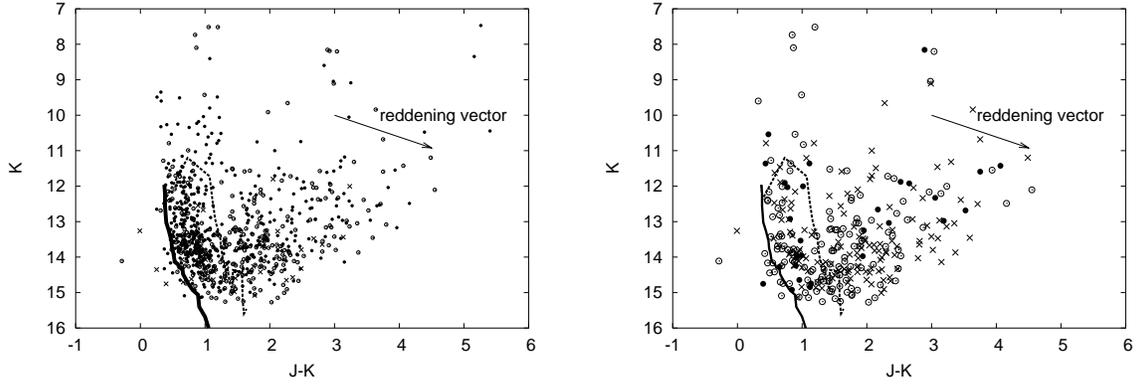}{Balog.fig7b.eps}
\caption{The J-K vs K diagrams of the detected IR sources (see Fig. 6 for the description of the plots.}
\end{center}
\end{figure}

\begin{figure}[ht]
\begin{center}
\leavevmode
\plottwo{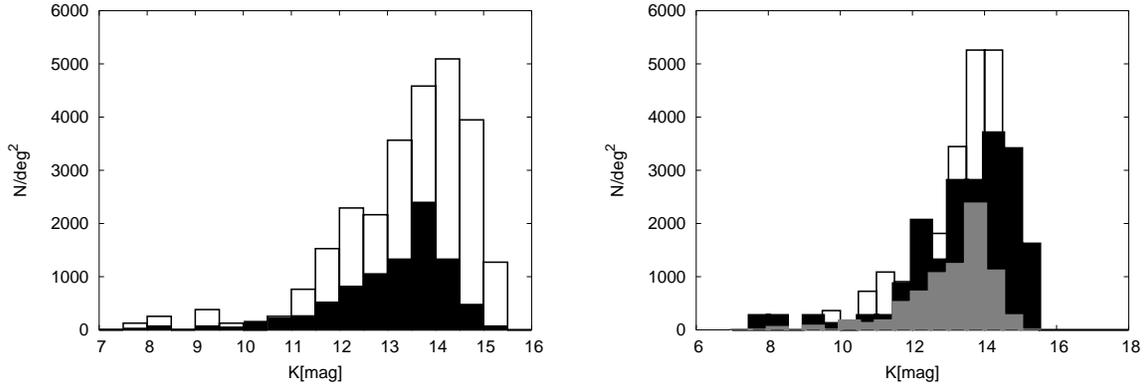}{Balog.fig8b.eps}
\caption{The $K$-band luminosity function. Left panel: all stars. Hollow histogram: concentration (within a 0.05 deg radius around the peak stellar density), filled histogram: remaining part of the area. Right panel: hollow histogram: region 2 of \citet{McCa91}, filled histogram: region 3 and grey filled histogram: remaining part of the area.}
\end{center}
\end{figure}

\begin{figure}[ht]
\begin{center}
\leavevmode
\plottwo{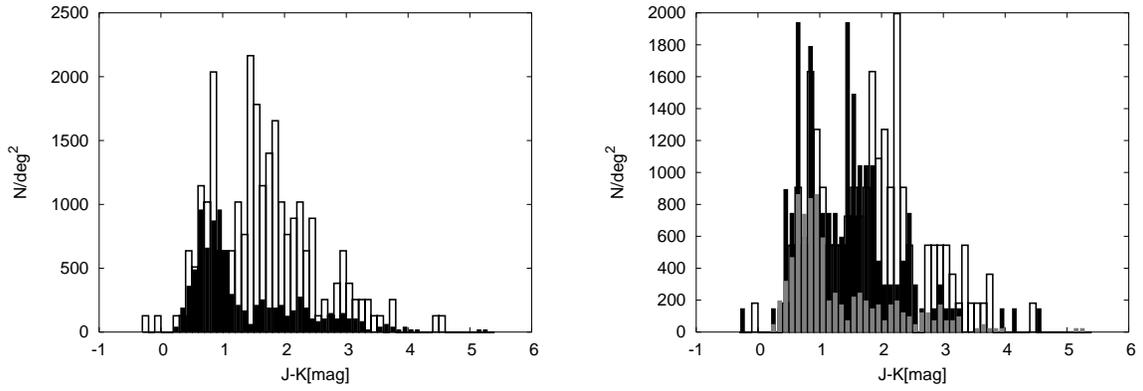}{Balog.fig9b.eps}
\caption{J-K color histograms of NGC~7538. See Fig. 8 for coding}
\end{center}
\end{figure}

\begin{figure}[ht]
\begin{center}
\leavevmode
\plotone{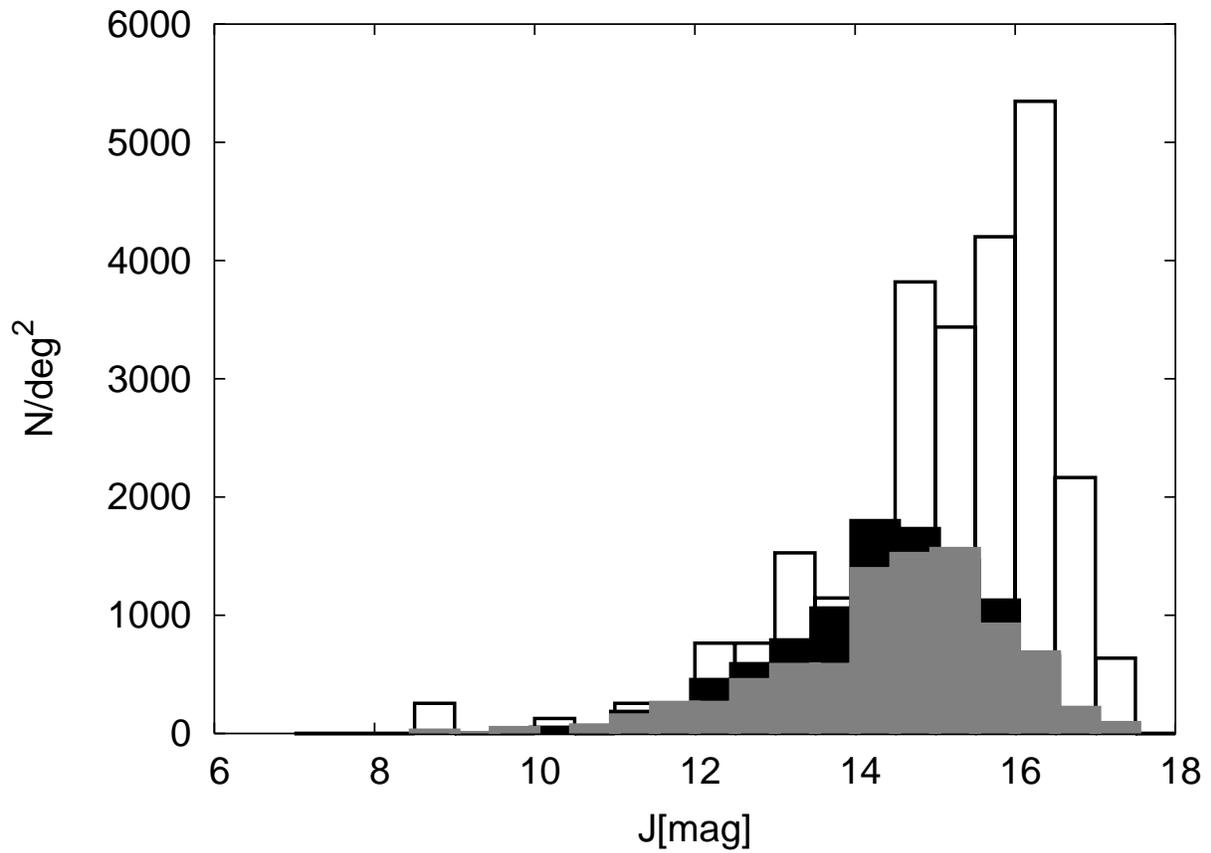}
\caption{J-band luminosity functions of NGC~7538. Hollow histogram: concentration as in Fig. 8 left panel; black filled histogram: stars outside the concentration; grey filled histogram: off-cluster control fields.}
\end{center}
\end{figure}

\begin{figure}[ht]
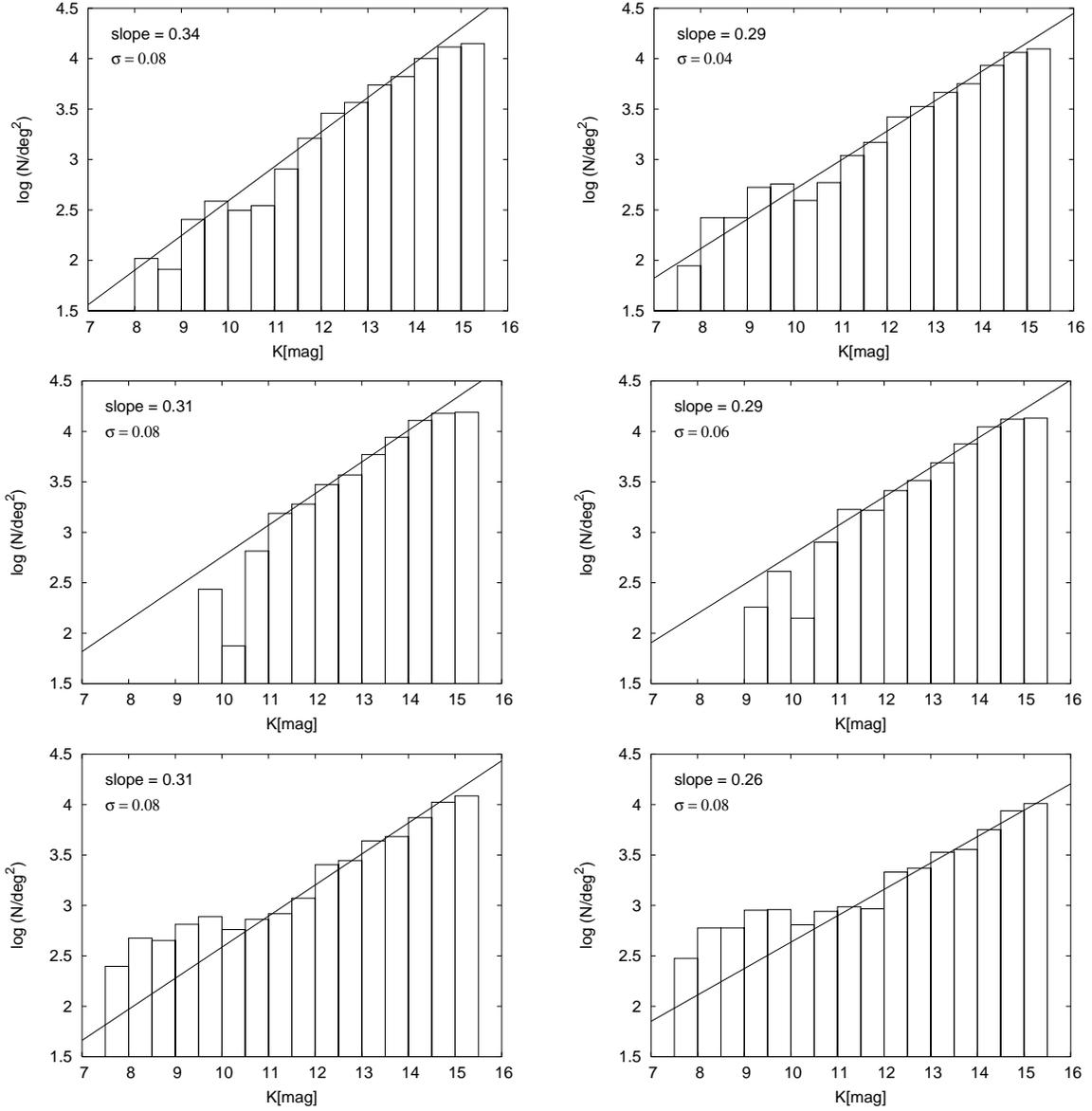

\begin{center}
\leavevmode
\plottwo{Balog.fig11a.eps}{Balog.fig11b.eps}
\plottwo{Balog.fig11c.eps}{Balog.fig11d.eps}
\plottwo{Balog.fig11e.eps}{Balog.fig11f.eps}
\caption{Cumulative logarithmic KLF of the concentration in NGC~7538. Top panels: all stars; middle panels: region 2 of \citet{McCa91}; bottom panels: region 3 of \citet{McCa91}. Left panels: KLF corrected with the on-cluster KLF; right panels: KLF corrected with the off-cluster KLF.}
\end{center}
\end{figure}

\end{document}